\documentclass[aps,preprint,nofootinbib,preprintnumbers,eqsecnum,superscriptaddress]{revtex4}

\usepackage{color}

\newcommand{\Comment}[1]{{}}
\usepackage[
      colorlinks=true,
      linkcolor=blue,
      urlcolor=blue,
      filecolor=black,
      citecolor=red,
      pdfstartview=FitV,
      pdftitle={},
        pdfauthor={Arpit Das, Nabil Iqbal, Napat Poovuttikul},
        pdfsubject={},
        pdfkeywords={},
        pdfpagemode=None,
        bookmarksopen=true
      ]{hyperref}

\usepackage[font=footnotesize,labelfont=bf,justification=centerlast,width=.94\textwidth]{caption}

\usepackage{threeparttable}

\usepackage[normalem]{ulem}
\usepackage{amsmath}
\usepackage{enumerate}
\usepackage{amsfonts}
\usepackage{yfonts}
\usepackage{array,xcolor,graphicx}

\usepackage{subfigure}
\usepackage{psfrag}

\usepackage{epsfig}
\usepackage[latin1]{inputenc}
\usepackage{float}
\usepackage{dsfont}
\usepackage{graphicx}
\usepackage{cancel}
\usepackage{mathrsfs}
\usepackage{amssymb}
\usepackage{amsfonts}
\usepackage{amsmath}
\usepackage{slashed}

\def\d{\partial}
\def\CO{\mathcal{O}}

\usepackage{graphicx}
\usepackage{bm}

\def\({\left(}
\def\){\right)}
\def\[{\left[}
\def\]{\right]}
\def\<{\langle}
\def\>{\rangle}

\def\CA{{\cal A}}
\def\CB{{\cal B}}

\def\CD{{\cal D}}
\def\CE{{\cal E}}

\def\CM{{\cal M}}

\def\CO{{\cal O}}

\def\CL{{\cal L}}

\def\CS{{\cal S}}

\newcommand{\bmat}{\begin{bmatrix}}
\newcommand{\emat}{\end{bmatrix}}

\newcommand\half{{\ensuremath{\frac{1}{2}}}}
\newcommand\p{\ensuremath{\partial}}

\newcommand{\be}{\begin{equation}}
\newcommand{\ee}{\end{equation}}
\newcommand{\bea}{\begin{eqnarray}}
\newcommand{\eea}{\end{eqnarray}}
\newcommand{\bwt}{\begin{widetext}}
\newcommand{\ewt}{\end{widetext}}

\newcommand{\bi}{\begin{itemize}}
\newcommand{\ei}{\end{itemize}}
\newcommand{\ben}{\begin{enumerate}}
\newcommand{\een}{\end{enumerate}}
\newcommand{\bca}{\begin{cases}}
\newcommand{\eca}{\end{cases}}
\newcommand{\bln}{\begin{align}}
\newcommand{\eln}{\end{align}}
\newcommand{\bst}{\begin{split}}
\newcommand{\est}{\end{split}}

\newcommand\ep{\epsilon}
\newcommand\sig{\sigma}
\newcommand\Sig{\Sigma}
\newcommand\lam{\lambda}
\newcommand\Lam{\Lambda}
\newcommand\om{\omega}

\newcommand\Ga{{\ensuremath{{\Gamma}}}}

\def\th{{\theta}}

\newcommand\ha{{\half}}

\def\le{\left}
\def\ri{\right}

\newcommand\sB{{\ensuremath{{\mathcal B}}}}

\newcommand\sL{{\ensuremath{{\mathcal L}}}}

\newcommand\sO{{\ensuremath{{\mathcal O}}}}

\newcommand{\NI}[1]{\textcolor{blue}{\textsf{[NI: #1]}}}
\newcommand{\AD}[1]{\textcolor{red}{\textsf{[AD: #1]}}}

\newcommand{\tG}{\tilde{G}}

\begin{document}

\title {Towards an effective action for chiral magnetohydrodynamics}

\author{Arpit Das}
	\email{arpit.das@durham.ac.uk}
	\affiliation{Centre for Particle Theory, Department of Mathematical Sciences, Durham University,
		South Road, Durham DH1 3LE, UK\\}
\author{Nabil Iqbal}
    \email{nabil.iqbal@durham.ac.uk}
    \affiliation{Centre for Particle Theory, Department of Mathematical Sciences, Durham University,
		South Road, Durham DH1 3LE, UK\\} 
\author{Napat Poovuttikul}
	\email{napat.po@chula.ac.th}
     \affiliation{Centre for Particle Theory, Department of Mathematical Sciences, Durham University,
		South Road, Durham DH1 3LE, UK\\} 
	\affiliation{Department of Physics, Faculty of Science
    Chulalongkorn University, Bangkok 10330, Thailand} 
\begin{abstract}
We consider chiral magnetohydrodynamics, i.e. a finite-temperature system where an axial $U(1)$ current is not conserved due to an Adler-Bell-Jackiw anomaly saturated by the dynamical operator $F_{\mu\nu} \tilde{F}^{\mu\nu}$. We express this anomaly in terms of the 1-form symmetry associated with magnetic flux conservation and study its realization at finite temperature. We present Euclidean generating functional and dissipative action approaches to the dynamics and reproduce some aspects of chiral MHD phenomenology from an effective theory viewpoint, including the chiral separation and magnetic effects. We also discuss the construction of non-invertible axial symmetry defect operators in our formalism. 
\end{abstract}

\vfill

\today
    \hypersetup{linkcolor=black}

\maketitle

\tableofcontents

\section{Introduction}
The framework of {\it hydrodynamics} describes the long-distance dynamics of conserved charges fluctuating around a state of thermal equilibrium \cite{landau1987fluid}. The structure of a hydrodynamic theory is (in principle) completely dictated by how the global symmetries of the system are realized in a state of thermal equilibrium. In general, we expect that given an understanding of the global symmetries, there should exist a precise algorithm to construct the corresponding effective hydrodynamic theory. 

In this work we discuss some progress towards formulating an effective theory of the chiral magnetohydrodynamic plasma. This can be understood as a finite-temperature system with a  $U(1)_{A}$ current that is not conserved due to the Adler-Bell-Jackiw anomaly \cite{Adler:1969gk,Bell:1969ts}, i.e. we have
\be
d\star j_A = -\frac{1}{4\pi^2} F\wedge F\,, \qquad \p_{\mu} j^{\mu}_A = -k \, \ep^{\mu\nu\rho\sig} F_{\mu\nu} F_{\rho\sig} \label{eq:WardIdenABJ}
\ee
where in this expression it is understood that we are examining a theory of {\it dynamical} electromagnetism, and that $F_{\mu\nu}$ is the field-strength tensor of a $U(1)$ gauge field. $k$ can be understood as a quantized anomaly coefficient; e.g. if we consider ordinary QED with massless Dirac fermions\footnote{A single species of Dirac fermions also has a $U(1)_{A}^3$ 't Hooft anomaly which we will ignore in this paper; it could be included by using the well-understood technology for the hydrodynamics of 't Hooft anomalies.} coupled to dynamical electromagnetism, we obtain this expression with $k = \frac{1}{16\pi^2}$. 

We stress that in this expression the right-hand side is a dynamical operator; the situation where the right-hand side is an external source (and where there need not be any dynamical gauge fields) is usually called a 't Hooft anomaly. The hydrodynamics of systems with 't Hooft anomalies is already a rich and well-understood field \cite{Son:2009tf, Neiman:2010zi}; see \cite{Landsteiner:2016led} for a review. 

The situation with an ABJ anomaly is at the moment somewhat less clear. Naively one might assume that as the current is not conserved, the corresponding symmetry is simply explicitly broken and plays no role in constraining the dynamics. This statement is somewhat too fast: indeed, recent work \cite{Choi:2022jqy,Cordova:2022ieu} has shown that in such a system, one can construct topological defect operators that count the axial charge, but these defect operators no longer obey a simple group composition law -- in other words the symmetry becomes {\it non-invertible} (a partial list of references on non-invertible symmetries in higher dimensions are \cite{Gaiotto:2019xmp,Koide:2021zxj,Choi:2021kmx,Kaidi:2021xfk,Heidenreich:2021xpr,Choi:2022zal,Roumpedakis:2022aik,Bhardwaj:2022yxj,Hayashi:2022fkw,Arias-Tamargo:2022nlf,Choi:2022fgx,Yokokura:2022alv}). This constitutes a precise non-perturbative characterization of the manner in which  the ABJ anomaly deforms the naive classical symmetry, and makes clear that -- at least in the vacuum -- a system with an ABJ anomaly is in a distinct universality class to one with no $U(1)$ symmetry at all. The understanding of the dynamical consequences of such a symmetry is still in its infancy; see e.g. \cite{GarciaEtxebarria:2022jky} for an extension of Goldstone's theorem to this setting and \cite{Karasik:2022kkq} for gauging of such non-invertible symmetries. 

The understanding of this non-invertible symmetry allows us to give a universal characterization of the chiral magnetohydrodynamic plasma: it is a system which realizes the non-invertible symmetry of \cite{Choi:2022jqy,Cordova:2022ieu} at finite temperature. In this work we make some attempts at describing the plasma from this point of view. 
%
%
The existence of such non-invertible defect implies that there exist a conserved 2-form current $J^{\mu\nu}$ and a 1-form current $j^\mu$ which satisfy the following Ward identity\footnote{Note that from here on we shall drop the subscript $A$ from the non-conserved 1-form current $j^\mu$ since we shall be studying a general effective theory which is in the same universality class as that of QED at finite temperature.}:
\be
\p_{\mu} J^{\mu\nu} = 0 \qquad \p_{\mu} j^{\mu} = k \, \ep^{\mu\nu\rho\sig} J_{\mu\nu} J_{\rho\sig} \label{symms} 
\ee
If the system admits a weakly coupled description in terms of a $U(1)$ photon whose field strength is $f_{\mu\nu}$, then $J^{\mu\nu} = \ha \ep^{\mu\nu\rho\sig} f_{\mu\nu}$; however we will not assume such a description in what follows. It is our understanding that any system which has vector and tensor fields that obey these two operator equations will allow the construction of the appropriate non-invertible defect operators; we will verify this in our constructions below. Our task is to understand how the symmetry structure \eqref{symms} is realized in thermal equilibrium (and for small fluctuations around it).

\subsection{Comparison to other approaches}\label{sec:compare}

The construction of the hydrodynamic description of a theory with an ABJ anomaly has a long history. For the convenience of the reader we briefly summarize some of this literature. 
 Early work in describing MHD in the presence of a finite chemical potential $\mu_A$ for the axial charge includes \cite{Joyce:1997uy}. They show the generic existence of instabilities in the presence of finite $\mu_A$.  
 
 Ref. \cite{Boyarsky:2015faa} constructs a description of the axial charge density in terms of an an effective dynamical axion $\th$.
They work in the limit where the fluid velocity is frozen $u^{\mu} = \delta^{\mu}_t$, and the presence of the dynamical axion means that the construction  somewhat resembles an axial superfluid; in particular their equations of motion depend on spatial gradients of $\th$.  A generalization of  \cite{Boyarsky:2015faa} to the case where the fluid velocity $u^{\mu}$ is dynamical can now be found in \cite{Rogachevskii:2017uyc}. Upon taking the limit of small magnetic diffusivity (i.e. that controls the diffusion of 1-form current), they found that $\partial_i\theta$ dependence drop out of the equations of motion. This set of equations is what is generally referred to as \textit{chiral magnetohydrodynamics}, and has numerous  astrophysics applications. A relativistic extension to a generic spacetime can be found in \cite{DelZanna:2018dyb}.

An interesting construction of dissipative chiral MHD from first principles using an entropy current was performed in  \cite{Hattori:2017usa}. The consistency of their derivative expansion required the anomaly coefficient $k$ in \eqref{eq:WardIdenABJ} to be a small parameter of order $\CO(\d^1)$. Another construction for the relativistic hydrodynamic description can be found in \cite{Yamamoto:2016xtu}, where the effect of the anomaly on the electric charge current \cite{Son:2009tf} is added to the Maxwell equation.  

Finally, an equilibrium effective action for the theory with 't Hooft anomaly organised with $u^\mu,T \sim \CO(\d^0)$ with the background gauge fields be $\CO(\d^0)$ and field strengths be $\CO(\d^1)$ can be found in e.g. \cite{Banerjee:2012iz,Jensen:2012jy,Jensen:2013vta}. In  \cite{Jensen:2013vta} the $U(1)$ current is weakly gauged, resulting in a system with an ABJ anomaly, and it was found that most transport coefficients receive radiative corrections. 

One major difference between our work and those earlier is that the bulk of the literature assumes the existence of a $U(1)_{V}^{(0)}$ vector electrical charge current, which is then coupled to dynamical electromagnetism in some manner, assuming some dynamics (e.g. Maxwell) for the electromagnetic sector. Philosophically, this can be thought of as weakly gauging a theory with a 't Hooft anomaly to convert it into an ABJ anomaly. 

From a modern point of view, however, the introduction of a photon which then interacts strongly with the plasma seems like an unnecessary intermediate step. An alternative approach is to attempt to bypass the weakly gauged construction completely, and simply directly attempt to describe the global symmetry structure, analogous to what was done for pure MHD in \cite{Grozdanov:2016tdf}. In this work we take some steps in that direction; i.e. we try construct an action-based approach to chiral MHD by realizing the symmetry structure \eqref{symms} directly in a minimal fashion without constructing an electric charge current or coupling it to a Maxwell field.  

Our work has something of an exploratory character; we do not write down the most general actions possible, rather constructing the simplest actions that display the required physics. We also always work in a limit where the fluid velocity and temperature are frozen.  Nevertheless, we will see that this is sufficient to reproduce many aspects of chiral MHD phenomenology.  

We now present a brief summary of the paper. In Section \ref{eqm}, we study the equilibrium sector of the hydrodynamic theory by placing the theory on $S^1\times\mathbb{R}^3$ and constructing an equilibrium generating functional. We study the decomposition of the symmetry breaking pattern under dimensional reduction and present an algorithm (order by order in the anomaly coefficient $k$) to compute the part of the action that is not invariant under gauge transformations of the axial source. We demonstrate that this construction leads to the chiral separation effect. 

After analysing the equilibrium sector, next we move onto the dissipative sector in Section \ref{diss}. We construct a real time effective action using the Schwinger-Keldysh formalism. We do this by ``gluing" together two independent theories for the 0-form and 1-form sectors in a way that preserves the anomaly structure, resulting in a dissipative action whose variation results in the expected equations of motion. A shortcoming with this construction (described in detail below) is that we are unable to preserve the so-called ``diagonal shift'' symmetry that is present in usual hydrodynamic actions describing 0-form symmetries in a normal phase. We conclude with a brief discussion in Section \ref{sec:conc}.  

\paragraph*{Note added:}  Towards the end of this work, we received \cite{Landry:2022}, which takes a different approach towards constructing an Schwinger-Keldysh effective action of the chiral plasma in Section \ref{diss} of this manuscript.

\section{Equilibrium sector}\label{eqm}
In this section, we develop an action which realises the finite-temperature equilibrium sector of a hydrodynamic theory which is in the same universality class as that of quantum electrodynamics (QED) at finite temperature. 
Since, this action would describe the equilibrium sector, it should not contain any time derivatives of the fields in it, by definition. We first review the symmetry structure of our theory. 

\subsection{Symmetries}
Consider a massless QED at finite temperature, whose weakly coupled physics is described by the following Lagrangian,
\begin{align}\label{weak_EM}
    \mathcal{S}[A,\Psi,\overline{\Psi}] = \int d^{4}x \left(- \, \frac{1}{g^2} F^2 + \overline{\Psi} \, \gamma^\mu\left(\partial_\mu - i A_\mu \right)\Psi\right),
\end{align}
where $A$ is the dynamical gauge field and $\Psi$ is a massless Dirac fermion and $F=dA$. The above action has a $U(1)^{(0)}$ axial current, denoted by $j^\mu = \overline{\Psi}\gamma^\mu\gamma^5\Psi$ which is non-conserved, due to the ABJ anomaly, and a $U(1)^{(1)}$ 2-form current, denoted by $J^{\mu\nu}=\frac{1}{2}\epsilon^{\mu\nu\rho\sigma}F_{\rho\sigma}$ which is conserved due to the Bianchi identity. Furthermore, the non-conservation of the axial current obeys the following equation (see appendix \ref{eqm_d_i} for details on how non-invertible defect insertion is equivalent to saturating the anomaly equation as in \eqref{eq:WardIdenABJ}),
\begin{align}\label{abj_anom}
	\partial_\mu j^\mu = k \, \epsilon_{\alpha\beta\rho\sigma} J^{\alpha\beta}J^{\rho\sigma}, \qquad \left(\text{where}, \, \, k\equiv\frac{1}{16\pi^2}\right).
\end{align}
Though we are inspired by massless QED, we will keep the constant $k$ arbitrary in what follows. The partition function is a function of two sources $a_\mu$ and $b_{\mu\nu}$ via 
\begin{equation}
    Z[a,b] = \int D[A]D[\Psi]D[\bar\Psi] \exp\left(-\CS + \int \left(\star j \wedge a + \star J \wedge b\right)\right)
\end{equation}
where $\epsilon^{\mu\nu\rho\sigma}(db)_{\nu\rho\sigma}\equiv j^\mu_\text{ext}$ can be thought of as an external current insert to the system. 

From here on, we will be agnostic to the details of the matter field except that it has the same global symmetry as the above theory. The equilibrium effective action we construct will be based only on the data encoded in the background field $a$ and $b$. This has the same spirit as \cite{Banerjee:2012iz,Jensen:2012jh,Armas:2018zbe} albeit in a more simplified metric and equations of state.
\subsubsection{1-form symmetry in thermal equilibrium}
Let us briefly review the notion of hydrodynamics with 1-form symmetries by considering ordinary MHD, i.e. a simple system in thermal equilibrium which has only a single 1-form symmetry (see \cite{Grozdanov:2016tdf, Iqbal:2020lrt}). The 1-form conservation equation takes the following form,
\begin{align}
    \partial_\mu J^{\mu\nu} = 0, \label{1-form_conserve}
\end{align}
Since we are interested in finite temperature physics, let us put our theory on $S^{1}\times\mathbb{R}^3$ and we shall denote the $S^1$ direction as $\tau$ (the Euclidean time). Now let us see how $J^{\mu\nu}$ decomposes in the dimensionally reduced theory on $S^{1}\times\mathbb{R}^3$. On $\mathbb{R}^3$ we have,
\begin{align}
    &U(1)^{(0)} \, \, \text{0-form symmetry} \, \, \to J^{i\tau} \to \mathcal{B}^i=J^{i\tau} \, \, \text{is magnetic 3-vector}, \label{decom_B} \\
    &U(1)^{(1)} \, \, \text{1-form symmetry} \, \, \to J^{ij} \to \mathcal{E}^i=\frac{1}{2}\epsilon^{ijk}J_{jk} \, \, \text{is electric 3-vector}. \label{decom_E}
\end{align}
Now, in equilibrium, to leading order in derivatives, $\mathcal{E}^i$ vanishes and the $U(1)^{(0)}$ symmetry is actually \textit{spontaneously broken} in the normal phase of the theory (see \cite{Iqbal:2020lrt}). So, for this spontaneously broken symmetry we will have a Goldstone mode which we denote by $\psi$ (this Goldstone mode may be thought of as the unscreened magnetic field in the plasma). Furthermore, due to the above symmetry breaking, $\psi$ has a shift symmetry of the form,
\begin{align}
    \psi\to\psi + \Lambda_{\tau}(x^i), \label{psi_transf}
\end{align}
where in the original theory, $\Lambda_\mu(x^i)$ is to be understood as a $\tau$-independent 1-form symmetry parameter. 

Now let us define the source for $U(1)^{(0)}$ to be $b_{i\tau}$. The transformation of $b_{i\tau}$ is as follows,
\begin{align}
    b_{i\tau}\to b_{i\tau} + \partial_i\Lambda_\tau, \label{bitau}
\end{align}
The source for $U(1)^{(1)}$ is going to be $b_{ij}$ with the following gauge transformation,
\begin{align}
    b_{ij}\to b_{ij} + \partial_i\Lambda_j - \partial_j\Lambda_i. \label{bij_gauge}
\end{align}
where $(i,j,k)=(x,y,z)$ to be considered in $\CM_3=\mathbb{R}^3$.

Let us now include the $0$-form non-conserved axial current $j^\mu$ in our theory. It decomposes on $S^1\times \mathbb{R}^3$ into $j^{\tau}$ and $j^i$. Note that the source for $j^{\tau}$ is $a_{\tau}$ and we define the source for $j^i$ to be $a_i$ which have the following gauge transformations,
\begin{align}
    &a_\tau\to a_\tau, \label{atau_g} \\
    &a_i\to a_i + \partial_i\lambda. \label{ai_g}
\end{align}
Note that due to the anomaly, the equilibrium action will ``strictly" not be gauge invariant with respect to the gauge transformations given in \ref{ai_g}, but this gauge non-invariance will be quite constrained by the anomaly equation given in \ref{abj_anom} as we shall see below.\footnote{Note that in the dimensionally reduced 3d theory we formally have a new non-invertible symmetry arising from the non-conservation of the current $j^{i}$: \be d \star j = k(\star J^{i\tau} \wedge \star J^{ij}), \ee where one views $J^{i\tau}$ and $J^{ij}$ as currents for $U(1)^{(0)}$ and $U(1)^{(1)}$ in the 3d theory respectively. Similar non-invertible symmetries have recently been studied in \cite{Damia:2022rxw}.} 

The gauge transformations \ref{psi_transf} and \ref{bitau} together indicate that $B_i \equiv \partial_i\psi-b_{i\tau}$ is a gauge-invariant 3-vector. Furthermore, we also have the following gauge-invariant tensors: $h_{ijk}\equiv\left(db\right)_{ijk}$ and $H_{ij}\equiv\left(dB\right)_{ij}$ and $\left(da\right)_{ij}$. These will be the basic building blocks for our hydrodynamic equilibrium action.

\subsection{Euclidean action}
Let us develop an effective action which describes the equilibrium sector of our theory. For this, let us first note that, in our theory $\psi$ is the only dynamical variable and $a_\mu$, $b_{\mu\nu}$ are sources for the currents $j^{\mu}$ and $J^{\mu\nu}$ respectively. Furthermore, let us note that this equilibrium effective action should be separately $C$, $P$, $T$, $CP$ and $CPT$ invariant since the microscopic theory in \ref{weak_EM} is invariant under each of these symmetries respectively. We tabulate in Table \ref{CPT} the transformation of each of the sources under the discrete symmetries (see Appendix \ref{app:discrete} for details). 

\begin{table}[h] 
\caption{Discrete Symmetry Table}
\label{CPT}
\begin{center}
\begin{threeparttable}
\begin{tabular}{l||c|c|c|c|c}
\hline
\hline
$\text{Symm.}$ & $a_{\tau}$ & $a_i$ & $b_{ij}$ & $b_{k\tau}$ & $\partial_i$ \\
\hline
\hline
P & $-1$ & $+1$ & $-1$ & $+1$ & $-1$ \\
T & $+1$ & $-1$ & $+1$ & $-1$ & $+1$ \\
C & $+1$ & $+1$ & $-1$ & $-1$ & $+1$ \\
CP & $-1$ & $+1$ & $+1$ & $-1$ & $-1$\\
\hline
\hline
\end{tabular} 
\end{threeparttable}
\end{center}
\end{table}

With this we can write down an effective action (upto $\mathcal{O}(\partial^2)$) for the equilibrium sector. We focus on the state where $a,b,B$ are small. The first few terms in this expansion that preserved all the discrete symmetries are as follow
\begin{align}
    \CS\left[\psi\right] = \int\limits_{\mathbb{R}^3}  & \left[\star\left(\frac{1}{2}\chi_A \left(a_{\tau}\right)^2\right) + \frac{1}{2}\chi_B\left(B\wedge \star B\right) + \, \frac{1}{2}\chi_C \, a_\tau \, \left(a\wedge da\right) + \frac{1}{2}\chi_G \, a_\tau \, \left(B\wedge dB\right) \right. \nonumber\\
    &\left.+ \frac{1}{2}\chi_I \left(H\wedge \star H\right) + \frac{1}{2}\chi_K \left(da\wedge \star da\right) + \, \frac{1}{2}\chi_N \, a_\tau \, (\star h) \, \left(B\wedge da\right) \right. \nonumber\\
    &\left. + \frac{1}{2}\chi_O (\star h) \, \left(h\right) + \, \frac{1}{2}\chi_P \, a_\tau \, (\star h) \, \left(a\wedge dB\right) \right], \label{second_a1}
\end{align}
The coefficient $\chi_A$ is the susceptibility of the axial charge sector and $\chi_B$ may be thought of as the susceptibility of the 1-form charge; physically it controls the amount of magnetic field produced in terms of a given 1-form chemical potential, which can be thought of as an applied external electric current \cite{Grozdanov:2016tdf}. 
 In the above action, each of the coefficients should be allowed to be an {\it even} function of $a_{\tau}$ (and an arbitrary function of the 0$^\text{th}$ order vector norm $B^2$). The extra explicit factors of $a_{\tau}$ render some of the coefficient functions odd under $a_{\tau} \to -a_{\tau}$ and guarantee the correct discrete transformation properties of the action. In the above action `$\star$' denotes the $3$-dimensional Hodge dual; we reserve the notation `$\star_4$' for the $4$-dimensional Hodge dual.

Finally, in this section we will seek to illustrate the minimum physics from imposing the anomaly constraint; let us then set all of the coefficients except $\chi_A,\chi_B,\chi_O$ to zero. Then we have,
\begin{align}
    \CS\left[\psi\right] = \int\limits_{\mathbb{R}^3} \, d^3x & \left[\frac{1}{2}\chi_A \left(a_{\tau}\right)^2 + \frac{1}{2}\chi_B\left(B_iB^i\right) + \frac{1}{12}\chi_O h_{ijk}h^{ijk} \right], \label{second_a2}
\end{align}
where we now further assume that the remaining $\chi_{A},\chi_B,\chi_O$ are simply constants. This action, however, does not have the non-invertible symmetry as the insertion of symmetry defect operator is not topological, see Appendix \ref{eqm_d_i} for further details. It turns out that the above action has to be modified by adding terms that are not gauge-invariant under \eqref{ai_g} in a very specific manner.
We will see that this is sufficient to reproduce some of the chiral MHD phenomenology. 
\subsubsection{Gauge non-invariant term} \label{sec:EquiCompat}
If the action above is gauge-invariant under transformations of the axial source \eqref{ai_g} and it will not yield the Ward identity in the dimensionally reduced theory i.e. 
\be\label{eq:WardDimReduced}
\partial_i j^i = k \, \epsilon_{ijk}J^{ij}J^{k\tau} \ .
\ee
To remedy this issue, we modify the action \eqref{second_a2} into the following form 
\begin{align}
    \CS\left[\psi\right] = \int\limits_{\mathbb{R}^3} \, d^3x & \left[\frac{1}{2}\chi_A \left(a_{\tau}\right)^2 + \frac{1}{2}\chi_B\left(B_i B^i\right) + \frac{1}{12}\chi_O \, h_{ijk}h^{ijk} + k \, a_i V^i \right], \label{ac_gni}
\end{align}
where the $V^i$ is an arbitrary vector that depends on $a_\tau, a^i, B^i$, their derivatives and $h=db$. This is due to the fact that the action is invariant under the background gauge transformation of $b_{ij}$ in \eqref{bij_gauge}. The currents in this theory can be written as 
\begin{subequations} \label{eq:constiReln-Equi}
\be\label{eq:constiReln-Equi-1}
\begin{aligned}
 j^\tau &= \frac{\delta \CS}{\delta a_\tau} = \chi_A \, a_\tau\,,  &\qquad
 j^i &=  \frac{\delta \CS}{\delta a_i} = k \, V^i + k \, \frac{\delta V^j}{\delta a_i}a_j, \,,\\
 J^{i\tau} &= \frac{\delta \CS}{\delta b_{i\tau}} = -\chi_B B^i + k \, \frac{\delta V^j}{\delta b_{i\tau}}a_j \, , &\qquad
 J^{ij} &= \frac{\delta \CS}{\delta b_{ij}}  = -\epsilon^{ijk}\partial_k f,
\end{aligned}
\ee
Notice that the constitutive relation for $J^{ij}$ is a total derivative of a 3-form $\mathcal{H}^{ijk} = \epsilon^{ijk} f$. This is due to the fact that $\CS$ can only depends on the the total derivative of $b_{ij}$. A pricise form of $f$ in terms of $V^i$  is 
\be \label{eq:constiReln-Equi-2}
f=\epsilon^{ijk}\left[\frac{\chi_O}{12}h_{ijk} + \frac{k}{6}a_m \, \frac{\partial \left(V^m\right)}{\partial (\partial_k b_{ij})}\right]\, . 
\ee
\end{subequations}

To find a form of $V^i$ which yield the Ward identity \eqref{eq:WardDimReduced}, we perform a transformation $a\to a+d\lambda$ in the action \eqref{ac_gni} to obtain the Ward identity 
\be\label{eq:WardFromac-gni}
\partial_i j^i = k \, \partial_i\left[V^i + \frac{\delta V^j}{\delta a_i}a_j\right]
\ee
Demanding the r.h.s. of \eqref{eq:WardFromac-gni} to be the same as those in \eqref{eq:WardDimReduced} and write $J^{ij}$ in terms of $f$ as in \eqref{eq:constiReln-Equi-1}, we find that 
\be\label{eq:Defect-consistency-cond}
\partial_i\left[V^i + \frac{\delta V^j}{\delta a_i}a_j\right] = \epsilon_{ijk} J^{ij} J^{k\tau} = - 2 \left[\d_i \left( J^{i\tau} f \right) - f \d_i J^{i\tau} \right] = -2 \d_i (J^{i\tau} f)
\ee
where, to get the last equality, we use the conservation law $\d_i J^{i\tau} =0$ in \eqref{1-form_conserve} (upon dimensionally reduced on the thermal cycle). 
Substitute the form of $J^{i\tau}$ and $f$ from \eqref{eq:constiReln-Equi-1} and \eqref{eq:constiReln-Equi-2} will provide us with a functional equation for $V^i$. 

Finding solutions to this is a well-posed but complicated task; while it seems possible that exact expressions should exist for arbitrary $k$ we have not been able to find them. To make progress, we thus consider a formal expansion in the anomaly coefficient $k$:  
\begin{equation}
\begin{aligned}
    V^i &= V^i_{(0)} + k \, V^i_{(1)} +\CO(k^2)\,,\\ 
\end{aligned}    
\end{equation}
Solving \eqref{eq:Defect-consistency-cond}, order by order in $k$, we find that 
\begin{equation}
    \begin{aligned}
V^i_{(0)} &= \frac{\chi_B\chi_O}{6} B^i\epsilon^{mpq}h_{mpq} = \chi_B\chi_O B^i |h|\,, \\
V^i_{(1)} &= V^i_{(1)} = \frac{\alpha}{2} a^i + \frac{\gamma}{2}\left(a\cdot B\right)B^i
    \end{aligned}
\end{equation}
where we denote $|h| = \frac{1}{6}\epsilon^{ijk}h_{ijk}$, $(a\cdot b)=a_i B^i$ , $\alpha \equiv (\chi_O^2 \chi_B)|h|^2$ and $\gamma \equiv (\chi_B^2\chi_0) $. 
Thus the action for non-invertible symmetry can be written as 
\begin{align}
\begin{split}
    \CS\left[\psi\right] &= \int\limits_{\mathbb{R}^3} \, d^3x \left[\frac{1}{2}\chi_A \left(a_{\tau}\right)^2 + \frac{1}{2}\chi_B\left(B_i B^i\right) + \frac{1}{12}\chi_O h^2\right.\\
    &\left. + k \, a_i \left\{\chi_B\chi_O |h| B^i + \frac{k}{2}\left(\chi_O \chi_B\right)\left[\chi_O |h|^2 a^i + \chi_B \left(a\cdot B\right)B^i\right]\right\}\right]
\end{split}
\end{align}
%

\subsubsection{Equations of motion}
We first derive the equations of motion for the $\psi$ field, $\frac{\delta \CS}{\delta \psi} = 0$,
\begin{align}
    \partial_l B_{l} + k \, \chi_O \partial_l \left(|h| a_l\right) = k^2 \, \chi_B\chi_O \partial_l\left[a_l \left(a\cdot B\right)\right]. \label{psi_eom}
\end{align}
We note the curious fact that due to the explicit presence of $a_i$ factor in the above equation, the equations of motion is no longer invariant under transformations of the axial source $a_i \to a_i + \p_i \lam$. 

Such phenomena occur even in simpler systems; for example let us consider axion electrodynamics, whose action takes the form:


\begin{align}
    \CS_{\text{axion}}[\theta,a] \sim \int \, d^4x \, \le[\left(d \theta - a\right)^2 + \theta F\wedge F + F^2 +  F \wedge b \ri] \label{axion_actn}
\end{align}
where $F=da$ and $b_{\rho\sigma}$ is the source for the 2-form current $J^{\mu\nu} \equiv \frac{1}{2} \epsilon^{\mu\nu\rho\sigma}F_{\rho\sigma}$. 

The equations of motion for $F$ from the above action are,
\begin{align}
    d\star F = Fd\theta, \label{eom_axion}
\end{align}
Clearly, under a shift of the axion, $\theta\to \theta + \Lambda$, the above equations of motion is no longer gauge-invariant, and it cannot be made so without spoiling gauge-invariance of the dynamical $U(1)$ gauge symmetry. This is arising from the existence of the gauge non-invariant term $\theta F\wedge F$ in the action; here we see a similar phenomenon, except that we do not even have a degree of freedom analogous to $\th$ in the action; rather the effect of the anomaly must be saturated using couplings to the axial source alone. 

It seems possible that there is a more elegant way to couple a source to this current so that some sense of invariance under transformations of the source is preserved, perhaps using the non-invertible symmetry structure. See \cite{Karasik:2022kkq} for some recent work in this direction. 


\subsubsection{Physical consequences}
In this section we interpret the result of the above section. Note that the magnetic field $\mathcal{B}$ is defined as, $\mathcal{B}^i = J^{i\tau}$ and the electric field $\mathcal{E}$ is defined as $\mathcal{E}_i=\frac{1}{2}\epsilon_{ipq}J^{pq}$. Now recall from \eqref{eq:constiReln-Equi-1} that the form of $J^{pq}$ is greatly constrained; as $J^{pq}=\frac{\delta\CS}{\delta b_{pq}} = -\partial_m\left(\mathcal{H}^{mpq}\right)$ for appropriately chosen $\mathcal{H}^{mpq}$, we see that $\mathcal{E}_i\sim\partial_i \Tilde{\alpha}$ where $\mathcal{H}^{mpq} = \Tilde{\alpha} \, \epsilon^{mpq}$. In other words, invariance under 1-form transformations of the source $b_{ij}$ \eqref{bij_gauge} is an extremely fancy way to demonstrate the elementary fact that in equilibrium, the electric field is always the spatial gradient of a potential. As is conventional, we will call this potential  the electric chemical potential $\mu_{\mathrm{el}}$. 

Putting in the specific expressions from above, we find (at $\mathcal{O}(k)$),
\Comment{
\begin{align}
    &\mathcal{E}_m^{(0)} = \p_{m} \mu_{\el} \label{elec0} \\
    \text{integrating,} \, \, &\mu_{\text{el}}^{(0)} = -\frac{1}{2}\chi_O\alpha_1, \label{muel0} \\
    &\mathcal{B}_m^{(0)} = -\chi_B B_m, \label{mag0}.
\end{align}
Using the above equations and \ref{anomaly_again} at $\mathcal{O}(k^0)$ we get,
\begin{align}
    j^i_A = 2k \, \mu_{\text{el}} \, \mathcal{B}^i, \label{cse0}
\end{align}
which is the \textit{chiral separation effect} (see Eq.(4.15) in \cite{Landsteiner:2016led}). 

Now let us analyse the situation at $\mathcal{O}(k)$. At this order we have,
}
\begin{align}
    &\mathcal{E}_m = \p_{m} \mu_{\mathrm{el}}, \label{elec1} \\
    \text{where,} \, \, &\mu_{\text{el}} \equiv -\frac{\chi_O}{2}\alpha_1 - \frac{k}{2}\chi_B\chi_O\left(a\cdot B\right), \label{muel1} \\
    &\mathcal{B}_m = -\chi_B B_m - k \, \chi_B\chi_O\alpha_1 a_m, \label{mag1}.
\end{align}
Note that in the absence of the anomaly, $\mu_{\mathrm{el}}$ is simply proportional to $h_{ijk}$; this is the component of the applied source that corresponds to applying an external electric charge density $\rho_{\mathrm{el}}$ \cite{Grozdanov:2016tdf}. 

Using the above equations and \eqref{eq:constiReln-Equi-1} we find (at $\mathcal{O}(k)$),
\begin{align}
    j^i = -2k J^{i\tau}f = k \left[\chi_O\chi_B\alpha_1 B^i + k\chi_O^2\chi_B a^i\alpha_1^2 + k \chi_O\chi_B^2\left(a\cdot B\right)B^i\right] = 2k\mu_{\text{el}}\mathcal{B}^i, \label{cse1}
\end{align}
In other words, up to the order in $k$ that we have been able to calculate, there is an axial current flow in the direction of the magnetic field, with coefficient precisely given by $k \mu_{\mathrm{el}}$. This is a well-known expression; it is the {\it chiral separation effect} \cite{Vilenkin:1980ft,Metlitski:2005pr,Newman:2005as}, see also \cite{Landsteiner:2016led} for a review. 

It is interesting to note that the coefficient of the chiral separation effect is given precisely by its value in the (ungauged) theory where we have a 't Hooft anomaly; the nonlinear terms in $k$ conspire precisely to make this possible. We stress that in our starting action we have chosen a minimal set of terms \eqref{second_a2} to explore the physics from the anomaly. It seems entirely possible that adding more terms will alter this relation, as is expected from \cite{Jensen:2013vta}, who found that including dynamical electromagnetism generically renormalizes all transport coefficients. We leave investigation of this important issue to later work. 

To conclude: here we constructed a Euclidean effective action that captures the simplest features of the axial anomaly. We demonstrated that it is possible (though somewhat cumbersome) to construct a generating function that saturates the anomaly equation; the resulting answers display the known physics of the chiral separation effect. We stress that we never weakly gauge electromagnetism; rather we simply directly discuss the universality class of the gauged theory.

\Comment{
\subsubsection{Saturating the anomaly}
\AD{Below will be changed based on modified action above...}
\begin{align}
    &d\star_4 j = k \, \star_4 J\wedge\star_4 J, \label{anomaly99} \\
    \text{leading to,} \, \, &\partial_i j^i = \frac{1}{4}\epsilon_{pqr\tau}J^{pq}J^{r\tau}, \label{anomaly9999} \\
    \text{leading to}, \, \, &\frac{1}{\beta}\left[\epsilon^{ijk}a_{[k,j]}a_{\tau,i}\chi_c + \epsilon_{ijk}\chi_z(a_\tau^2 h_{mij} B_k)_{,m}\right] \nonumber\\ 
    &= \frac{1}{4}\chi_O\chi_g(a_\tau^2h^{mjk})_{,m}(B_j a_\tau)_{,k} - \frac{1}{4}\chi_O\chi_g(a_\tau^2h^{mjk})_{,m}B_{[j,k]} a_\tau \nonumber\\
    &+ \frac{1}{8}\chi_O\chi_B(a_\tau^2h^{mjk})_{,m} B_i\epsilon_{jki} \label{anomaly_contraint_0}
\end{align}
To find non-trivial relations let us simplify \ref{anomaly_contraint_0} by putting $\chi_c=0=\chi_g$. Then we get (at $\mathcal{O}(\partial^2)$),
\begin{align}
    &\chi_z = \frac{k\beta}{8}\chi_O\chi_B. \label{ac1}
\end{align}
To obtain the above relation we have also put $\psi$ on-shell which means $B_{i,i}=0$. Thus, \ref{ac1} is the non-trivial constraint on transport coefficient we get from saturating the anomaly.
}


\section{Dissipative action}\label{diss}

In this section, we move beyond the equilibrium construction of the previous section and construct a dissipative action that realizes this anomaly structure. We will use the recently constructed formalism for finite-temperature dissipative actions \cite{Crossley:2015evo,Glorioso:2017fpd,Haehl:2014zda}; see \cite{Liu:2018kfw} for a review of this technology. Here we assume the reader already has some familiarity with that formalism. As we freeze the stress-energy sector we will require only a subset of the full technology. We will construct an effective action representing the symmetry structure described above, and we will see that already interesting physics appears at the first order in derivatives.

We note from the outset that while this does result in a useful action principle for obtaining the correct equations of motion, we do not currently feel that this is the most elegant formulation of the problem, for reasons explained below. 

The strategy that we take is to first consider two theories which capture the dissipative dynamics of the charges associated with a conserved $U(1)$ 0-form symmetry (with an associated axial charge current $j^{\mu}$ ) and a 1-form symmetry (with an associated magnetic flux current $J^{\mu\nu}$). We then ``glue'' these two theories together in a manner which results in the $U(1)$ 0-form symmetry being broken down in the manner described by the anomaly equation \eqref{symms}. We will see that the implementation of this structure is most convenient if we introduce some auxiliary fields; upon eliminating these fields we find \eqref{symms} on-shell. 

We begin with the Lagrangian densities for the two subsectors for the 1-form and 0-form sectors respectively: 

\begin{align}
\sL_{0}[a;\th] & = \frac{i\sig}{\beta} A_{ai}^2 + \chi_A A_{a0}A_{r0} - \sigma A_{ai} \p_{0} A_{ri} \label{j0} \\
\sL_{1}[b;\Phi] & = \frac{i \rho}{\beta}G_{aij}^2 + \chi_B G_{a 0i}G_{r 0i} - \rho G_{aij} \p_{0} G_{rij} \label{mhd} 
\end{align}
where we have
\be
A = a + d\th \ , \qquad G = b + d\Phi\,.
\ee
where the time-component of the 1-form field $\Phi_t$ reduced to $\psi$ in equilibrium configuration presented in Section \ref{eqm}. The Lagrangian \eqref{j0} and \eqref{mhd} each describes the charge diffusion process of 0-form and 1-form symmetry respectively. It is also follows that this is the most general Lagrangian in each sector that preserved all C, P and T symmetry at first order in derivative and quadratic order in amplitude of $A$ and $G$ see e.g. \cite{Liu:2018kfw,Vardhan:2022wxz}.

Here and throughout we use a notation where lowercase letters are applied sources and uppercase or greek letters are dynamical fields. For illustrative purposes we work with the simplest possible actions, i.e. only keeping terms to quadratic order in the fields. \eqref{j0} describes the diffusive dynamics of an ordinary 0-form conserved charge in terms of the scalar Stuckelberg $\th$, the construction is reviewed in \cite{Liu:2018kfw}. \eqref{mhd} has recently been constructed in \cite{Vardhan:2022wxz} to describe diffusive dynamics of the magnetic field in terms of a vector Stuckelberg $\Phi$; it should be clear that it is a 1-form generalization of \eqref{j0}. 

In general one obtains the currents through functional differentiation with respect to the sources:
\be
j^{\mu}_{r,a} = \frac{ \delta \sL}{\delta a_{\mu}^{a,r}} \qquad J^{\mu\nu}_{r,a} = \frac{\delta \sL}{\delta b_{\mu\nu}^{a,r}} \label{currdef} \ . 
\ee
As usual, the invariance of the action under the following combined transformations of the sources and dynamical fields:
\begin{align} 
a \to a + d\Lam(x) \qquad \th \to \th - \Lam(x) \\
b \to b + d\xi(x) \qquad \Phi \to \Phi - \xi(x) 
\end{align} 
(where $\Lam$ is a scalar and $\xi$ a 1-form) implies conservation of the currents $j^{\mu}$ and $J^{\mu\nu}$ as defined in \eqref{currdef}. 

Let us briefly discuss the physical interpretation of the coefficients appearing in the action above. $\rho$ is the electrical resistivity; as stressed in \cite{Grozdanov:2016tdf}, in a universal formulation of magnetohydrodynamics, it is $\rho$ that is a fundamental transport coefficient, and not the electrical conductivity. In particular, $\rho$ can be matched to microscopics through the following Kubo formula, in terms of the retarded correlator of the $2$-form current $J^{xy}$. 
\be
\rho = \lim_{\om \to 0} \frac{1}{-i\om} G^{R}_{xy,xy}(\om) \label{rhodef} 
\ee
$\sigma$ is the conductivity of the $U(1)$ 0-form axial charge (and is unrelated to the vector electrical conductivity). $\chi_A$ and $\chi_B$ are charge susceptibility of 0-form and 1-form symmetry that appears in the zeroth derivative level of \eqref{second_a1}.  

\subsection{Combining theories using auxiliary fields}
We would now like to glue these two theories together so that $j^{\mu}$ is no longer precisely conserved, but instead so that we find the following expression in the final combined theory:
\be
\p_{\mu} j^{\mu} = k \ep^{\mu\nu\rho\sig} J_{\mu\nu} J_{\rho\sig}  \label{Janom} 
\ee
where this expression is now understood to hold on both legs of the doubled Schwinger-Keldysh contour, i.e. for the 1-type and 2-type fields individually. 
To do so, we introduce two new sets of auxiliary fields: two 2-forms $\Sig^{r,a}$ and two 1-forms $C^{r,a}$. These fields are useful to ``unwrap'' the non-linearities that are present in the anomaly equation; to obtain the physical currents they should be eliminated, as we do explicitly below. We thus consider the following combined action:
\be
\sL[a,b;\th,\Phi,\Sigma,C] = \sL_0[a;\th] + \sL_{1}[b + \Sig; \Phi] - \ha\le(\ep^{\mu\nu\rho\sig} \Sig^a_{\mu\nu} dC^r_{\rho\sig} + \ep^{\mu\nu\rho\sig} \Sig^r_{\mu\nu}  dC^a_{\rho\sig}\ri) + \sL_{anom}[\th,C], \label{diss_actn}
\ee
where $\sL_{anom}[\th,C]$ takes the form:
\be
\sL_{anom}[\th,C] = -k \le(\th_{1} \ep^{\mu\nu\rho\sig} dC^1_{\mu\nu} dC^1_{\rho\sig} - \th_2 \ep^{\mu\nu\rho\sig} dC^2_{\mu\nu}dC^2_{\rho\sig} \ri)
\ee
Note the direct coupling to the Stuckelberg field $\theta$ clearly breaks its shift symmetry. It will often be useful for us to rewrite this action in the $r-a$ basis:
\be
\sL_{anom}[\th, C] = k \le(\th^a \le(\ep^{\mu\nu\rho\sig} dC^r_{\mu\nu} dC^r_{\rho\sig} + \frac{1}{4}\ep^{\mu\nu\rho\sig} dC^a_{\mu\nu} dC^a_{\rho\sig}\ri) + 2 \th^r \ep^{\mu\nu\rho\sig} dC^a_{\mu\nu} dC^r_{\rho\sig}\ri) \label{anombit} 
\ee
Varying the action with respect to $\th^{1,2}$, we find:
\be
\p_{\mu} j^{\mu}_1 = - k \ep^{\mu\nu\rho\sig} dC^1_{\mu\nu} dC^1_{\rho\sig} \label{jvarC} 
\ee
We now note that the $\Sigma$ fields couple to the 1-form sector as a direct shift of the external 2-form source $b$, i.e. always in the combination $b + \Sigma$. Thus the equations of motion for the auxiliary field $\Sig^{r,a}$ are determined by the variation of $\sL_{1}$ with respect to $b$: $\delta_{\Sig}\sL_{1} = \delta_{b}\sL_{1}$. This results in the following equations of motion:
\be
\frac{\delta \sL_{1}}{\delta b^{a}_{\mu\nu}} = \ha \ep^{\mu\nu\rho\sig}dC^{r} \label{onshellval} 
\ee
(and similarly for $r \leftrightarrow a$). However the left-hand side is by construction the 2-form current $J^{r}$ \eqref{currdef}. Thus we see that the role of the $\Sigma$ fields is to simply to precisely correlate the $C^{r,a}$ fields with the 2-form currents as:
\be
J^{\mu\nu}_{r,a} =  \ha \ep^{\mu\nu\rho\sig}dC^{r,a}_{\rho\sig} \label{Ceq} 
\ee
Inserting this into \eqref{jvarC} we find exactly the desired expression \eqref{Janom}; thus this construction always relates the non-conservation of the axial current with the magnetic flux in the correct fashion. 

A modern understanding of \eqref{Janom} is that it permits the construction of topological defect operators that measure the axial charge \cite{Choi:2022jqy,Cordova:2022ieu}; in Appendix \ref{app:defectop} we verify that such defect operators can be constructed in this theory (indeed we have one such operator living on each of the legs of the Schwinger-Keldysh contour). 

We note some facts about this construction:
\ben
\item The structure does not depend on the precise form of the 0-form and 1-form theories $\sL_{0}$ and $\sL_{1}$, but only on their invariances under symmetries. 
\item As none of the new terms we have added introduce any couplings between the two legs of the time contour, the action is automatically invariant under the so-called {\it KMS symmetry}; this acts on all fields and sources $\phi$ as
\begin{align}
\phi_{a}(x) & \to \Theta \phi_{a}(x) + i \beta \Theta \p_t \phi_{r}(x) \\
\phi_{r}(x) & \to \Theta \phi_{r}(x) 
\end{align}
with $\Theta$ an anti-unitary symmetry representing time-reversal.
\item The action has a rather undesirable feature: it is not invariant under the ``diagonal shift'' symmetry in the scalar sector. To be more precise, in an action-based formulation of hydrodynamics one typically requires that the action be invariant under shifting the $r$-type Stuckelberg $\th^r$ by an arbitrary spatially dependent phase, i.e.
\be
\th^{r}(t,x^i) \to \th^{r}(t,x^i) + \lam(x^i) \label{diagshift} 
\ee
where $\lam(x^i)$ is an arbitrary function of space \cite{Dubovsky:2011sj,Crossley:2015evo,Glorioso:2017fpd}. Invariance under this symmetry is generally required to forbid arbitrary spatial gradients of $\p_i \th^r$ in the action or constitutive relations; if this diagonal symmetry is broken then one should add such spatial gradients to the action, and we then generically end up in a superfluid phase for the corresponding symmetry (in this case $U(1)_{A}$). 

In our construction, the non-anomalous part of the action \eqref{j0} is invariant under the ``diagonal shift'' symmetry, but the anomalous part \eqref{anombit} is {\it not}. In the dual variables that we are using, it is not straightforward to make the action invariant under this diagonal shift. This is undesirable: in the case of the 't Hooft anomaly, the interplay between the diagonal shift and the realization of the anomaly plays a very important role \cite{Glorioso:2017lcn,Haehl:2013hoa,Dubovsky:2011sk}. 
\een
At the moment, we are unclear on the precise implications of breaking this diagonal shift symmetry. We will proceed with this action and find physically very reasonable results; however, as there is no symmetry preventing us from adding $\p_{i} \th^r$ terms to the action we cannot in good-faith call this an {\it effective} field theory; rather it is simply an action which one can use to obtain a consistent set of equations of motion. Given the unclear formal status of chiral MHD, this still seems to be of value, and we leave to the future a more refined understanding of the interplay of the diagonal shift symmetry and the non-invertible character of the axial anomaly.  

\subsection{Chiral MHD phenomenology}\label{sec:chiralMHDphenomenology}
We now study some simple consequences of varying this action. From \eqref{currdef} above we have from the magnetic sector
\begin{align}
J^{0i}_{r} = \chi_B \tG_{r0i} \qquad J^{ij}_r & = \frac{2i \rho}{\beta} \tG_{aij} - \rho \p_0 \tG_{rij} \label{Jdef} \\
J^{0i}_{a} = \chi_B \tG_{a0i} \qquad J^{ij}_a & = \rho \p_{0} \tG_{aij}
\end{align} 
where we have defined a shifted $G$ which takes into account fluctuations of the new auxiliary field $\Sigma$: 
\be
\tilde{G} \equiv G + \Sigma = b + d\Phi + \Sigma \ . 
\ee
Similarly, in the axial charge sector we have:
\begin{align}
j^0_{r} = \chi_A A_{r0} \qquad j^i_{r} & = \frac{2i\sig}{\beta} A_{ai} - \sig \p_{0} A_{ri} \\
j^0_{a} = \chi_A A_{a0} \qquad j^i_{a} & = - \sig \p_{0} A_{ai}
\end{align}
Finally, varying the action with respect to $C^{a}$, we find the following expression for $\Sig$:
\be
\p_{[\rho} \Sig^{r}_{\mu\nu]} = 4 k \p_{[\rho} \th^{r} dC^{r}_{\mu\nu]} \label{sigeq} 
\ee
When studying classical equations of motion it is self-consistent to set all $a$-type fields to zero after variation of the action; we have done this above. In the remainder of this section we will thus omit the ``r'' superscript on all quantities; everything that remains is an $r$-type field. 

As usual, we now define the axial chemical potential to be
\be
\mu_{A} = A_{0} = \le(\p_{0}\th + a_{0}\ri)
\ee
It is also convenient to define the following vector ``chemical potential'' for the 1-form charge:
\be
\mu_i = \tG_{0i} = (b + \Sigma + dA)_{0i}
\ee
so that we can write
\be
\p_{0}(b + \Sig + dA)_{ij} = h_{0ij} + (d\Sigma)_{0ij} + \p_{i}\mu_j - \p_{j}\mu_i 
\ee
where $h = db$. We can then write the currents as
\begin{align}
J^{0i} & = \chi_B \mu^i \\
J^{ij} & = -\rho(\p_{i}\mu_j - \p_{j}\mu_i + (d\Sigma)_{0ij}) \label{Jsigmarel} 
\end{align}
where we have set the sources $h = db = 0$.

We now see the first effect of the anomaly; the $\Sigma$ field is now contributing to the spatial components $J^{ij}$; in a conventional formulation of the theory this component of the 2-form current is the electric field. We may explicitly find expressions for the currents by using \eqref{sigeq} and \eqref{Ceq} to eliminate $dC$ and $\Sigma$. To this order in derivatives this is a linear set of equations that can in principle be straightforwardly solved; in practice the expressions are somewhat cumbersome. 

We present first the answer assuming the the system is spatially homogenous ($\p_{i} = 0$). We set to zero all sources except for the axial source $a_t$. We then find:
\be
(d\Sig)_{0ij} = 4 k (\p_{0} \th) \ep_{ijk} J^{0k}
\ee
which then leads to the following expressions for the currents
\begin{equation}\label{eq:constitutiveReln}
\begin{aligned}
j^{0} & = \chi_A \mu_A \qquad j^i = 0 \\
J^{0i} & = \chi_B \mu^i  \qquad J^{ij} = -2k \rho \le(\mu_A - a_{t}\ri) \ep^{ijk} J^{0k}
\end{aligned}
\end{equation}
Let us examine the expression for $J^{ij}$; we see that the same transport coefficient $\rho$ that determines the resistivity determines the strength of the electric field $\mathcal{E}^{i} \sim k\rho(\mu_A - a_t) \mathcal{B}^i$. This is a manifestation of the chiral magnetic effect. In a more conventional weakly-gauged description, this arises from considering the vector current $j_{\mathrm{el}}^i \sim k \mu_A B^i$ and relating it to the electric field through the electrical conductivity $\mathcal{E}^i = \sigma_{\mathrm{el}} j_{\mathrm{el}}^i$. However, in a formulation of MHD based only around symmetries, it is difficult to give a precise meaning to either $j^{\mathrm{el}}$ or $\sigma_{\mathrm{el}}$ \cite{Grozdanov:2016tdf}; here we see  (as expected) that in this dual language the CME is controlled by $\rho$ instead.

The equations of motion arise from varying the action with respect to $\th$ and $\Phi$, and are the (non)-conservation of the respective currents:
\be
\p_{\mu} j^{\mu} = k \ep^{\mu\nu\rho\sig} J_{\mu\nu} J_{\rho\sig} \qquad \p_{\mu} J^{\mu\nu} = 0 \label{eomf} 
\ee
Putting in the constitutive relations for the currents, we now find that $\mu_i$ is constant in time, but that $\mu_A$ necessarily evolves according to the following equation:
\be
\dot{\mu}_A =  -2 k^2 \frac{\rho}{\chi_A} (J^{0i})^2 (\mu_A - a_t)
\ee
Thus we find that $\mu_A$ relaxes towards the externally applied $a_t$ with an exponential decay, i.e.
\be
\mu_A - a_t \sim e^{-\Ga_{A} t} \qquad \Ga_A = 2 k^2 \frac{\rho}{\chi_A} (J^{0i})^2 \label{disprate} \,.
\ee
This equilibrium configuration is consistent with what one would get from gauging procedure shown in \eqref{sec:EqFromGauging}.
It also allows us to make sense of the constitutive relation in Eq. \eqref{eq:constitutiveReln} as a small expansion around equilibrium configuration in the late time limit where $e^{-\Gamma_A t}\ll 1$. This kind of procedure is common in the study of hydrodynamics with weakly broken global symmetry, see e.g. \cite{Grozdanov:2018fic,Stephanov:2017ghc} for recent discussions.
The presentation of the decay rate $\Gamma_{A}$ suggests a useful formula for it in the small $J^{0i}$ limit, i.e.
\be
\Gamma_{A} = c (J^{0i})^2 \qquad c = \frac{2 k^2}{\chi_A} \le( \lim_{\om \to 0} \frac{1}{-i\om} G^{R}_{xy,xy}(\om)\ri) \label{kubogam} 
\ee
where we have used the Kubo formula for $\rho$ \eqref{rhodef}. 

 We see that as $t \to \infty$, an equilibrium configuration can have a nonzero value of $J^{0i}$ (as dictated by the unbroken 1-form symmetry), but that $\mu_A$ will always be equal to $a_t$, and that even homogenous fluctuations around this value are damped. Of course for an ordinary conserved current fluctuations of $\mu$ obey a diffusion equation and are undamped in the homogenous limit.

\subsection{Spatial derivatives}

We now allow nonzero spatial dependence, i.e. we allow $\p_i \neq 0$. It was demonstrated above that the equilibrium state takes the form:
\be
\mu_A = a_t \qquad J^{0i} = \chi_B \mu^i
\ee
 where all the fields are constant in space and time. We now consider linearized perturbations $\mu^i \to \mu^i + \delta\mu^i$, $\mu_A \to \mu_A + \delta\mu_A$ around this background configuration. We find the following expressions for the currents: 
\begin{align}
J^{0i} & = \chi_B(\mu^i + \delta\mu^i) & \qquad
 J^{ij} & = -\rho \le(\p_{i} \delta \mu_j - \p_{j} \delta \mu_i\ri) - 2k \rho \chi_B \delta \mu_A \ep^{ijk} \mu_k \\
j^0 & = \chi_A \delta \mu_A  & \qquad j^{i} & = -\sig \p_i \delta\mu_A
\end{align}
where as before we have assumed that the external sources have vanishing field strength, i.e. $da = db = 0$. 

These expressions are what one might expect on physical grounds; in particular we record the expression for $J^{ij}$ in the conventional language of the electric field $\CE^i$:
\be
\mathcal{E} = \rho(\chi_B^{-1} \nabla \times \mathcal{B} - 2 k (\mu_A - a_t) \mathcal{B}) \label{finalac}
\ee
(where the first term receives contributions only from the fluctuations in $\mathcal{B}$, whereas the second receives contributions from fluctuations in $\mu_A$ and the background in $\mathcal{B}$). 

We note that the equations of motion can be written only in terms of $\mu_A$, and do not require explicit mention of the Stuckelberg field $\th$. This happens because $\th$ enters only through $\p_t \th$, and spatial gradients $\p_i\th$ do not appear; this can be traced back to the precise form of the expression \eqref{sigeq} and \eqref{Jsigmarel}. We note that this was not obviously guaranteed. In the usual formulation of effective actions for hydrodynamics this property is enforced by the diagonal shift symmetry. As we noted previously, our system does not have this symmetry, and it seems possible that at higher orders in non-linearities and/or derivatives spatial gradients of the Stuckelberg field will appear. We leave this possibility for later investigation. 

As a simple application we study the dispersion relations in this framework. We orient the background field in the $z$ direction $\mu^z$ and consider a fluctuation $\delta \mu^y$ with momentum $q$ in the $x$ direction\footnote{Had we set $q$ to be in the direction parallel with $\mu^i$, the two modes modes $\omega_1,\omega_2$ depends quadratically on $q$ i.e.
\begin{equation}
    \omega_1 = -\frac{2i \CB^2 k^2 \rho}{\chi_A} - iq^2\frac{\sigma }{\chi_A} \, ,\qquad \omega_2 = -iq^2 \frac{\rho}{\chi_B}\,.
\end{equation}
}, so that perturbations have the spacetime dependence $e^{-i\om t + i q x}$. From \eqref{eomf} it is straightforward to find two modes $\om_{1,2}(q)$. The expressions are somewhat cumbersome, so we record them in two limits. We assume $\frac{\rho}{\chi_B} < \frac{\sig}{\chi_A}$, and we first present the ``high'' momentum limit:
\begin{align} 
\om_1(q) & = -iq^2\frac{\sig}{\chi_A}  \le(1 - \frac{2 k^2 \sB^2 \chi_B}{\rho \chi_A - \sig \chi_B} \frac{1}{q^2} + \sO(q^{-4})\ri) \\
\om_2(q) & = -iq^2 \frac{\rho}{\chi_B} \le(1 + \frac{2 k^2 \sB^2 \chi_B}{\rho \chi_A - \sig \chi_B} \frac{1}{q^2} + \sO(q^{-4})\ri) 
\end{align}

Here ``high'' means that $q^2 \gg \frac{k^2 \sB^2 }{\sig}$, i.e we are looking at scales higher than the scale determined by the anomaly. We see that in this regime the two modes are essentially those of the diffusion of conventional 0-form charge and that of magnetic field lines respectively, with the two diffusion constants set by $D_{a} = \frac{\sig}{\chi_A}$ and $D_{b} = \frac{\rho}{\chi_B}$; i.e. in this regime we find the physics of the original non-anomalous model.  

At low $q$ we find instead the following interesting dispersion relations, which we expand in the first few orders in the spatial momentum $q$. 
\begin{align} 
\om_1(q) & = \frac{-2 i \sB ^2 k^2 \rho}{\chi_A} - i q^2 \le(\frac{\rho}{\chi_B} +  \frac{\sig}{\chi_A}\ri) + \sO(q^4) \\
\om_2(q) & = -i\frac{\sig}{2 \sB^2 k^2 \chi_B}q^4 + \sO(q^6)
\end{align}
We see that the 0-form diffusive mode is now gapped in the IR, as we saw above in \eqref{disprate}.  The leading momentum-dependence of this mode now has a dissipative character, where the dissipation rate is interestingly controlled by the {\it sum} of the diffusion rates of the original magnetic and axial charge sector respectively. 

The second ``subdiffusive'' mode -- which at high momenta becomes the diffusion of magnetic flux --interestingly starts at $\sO(q^4)$ unlike the usual diffusion where $\omega \sim -i q^2$ commonly found in hydrodynamic systems. It should be noted here that the modes $\omega \sim -iq^4$ has been observed in various anisotropic systems with intricate symmetry structure such as a system with 't Hooft anomaly at strong mangetic field \cite{Ammon:2020rvg}, systems with conserved multipole moment \cite{Gromov:2020yoc,iaconis2021multipole} and easy-axis Heisenberg spin chain with integrability-breaking perturbation \cite{de2022subdiffusive} to name a few.  
Its physical origin seems to be tied to these somewhat exotic symmetry patterns and deserves further investigation.

We should also mentioned that, often in the chiral MHD literature, the system exhibits various kinds of interesting instabilities which have potential applications in astrophysical plasma. However, our focus is on the perturbation around equilibrium configuration and study how the system relaxed back to equilibrium. Had we chosen to perform a perturbation around constant $\mu \ne 0$ and $a_t=0$, we will also find unstable solution as those in e.g. \cite{Joyce:1997uy}.
\section{Discussion and outlook} \label{sec:conc} 
In this work we documented some progress towards a description of chiral MHD that relies only on the global symmetries. One of our results was be an expression for the low-field limit of the chiral charge relaxation rate $\Gamma_{A}$. Our work suggests that in the limit of small $\mathcal{B}$ field, this expression is universal, and takes the form:
\be
\Gamma_{A} = c \mathcal{\sB}^2 \qquad c = \frac{2k^2}{\chi_A} \lim_{\om \to 0} \frac{1}{-i\om} G^{R}_{\mathcal{E}^z,\mathcal{E}^z}(k \to 0, \om)
\ee
where we have rephrased \eqref{disprate} in terms of conventional electric and magnetic fields. 

Formulas of this sort are well-known (see e.g. \cite{Das:2022auy,figueroa1}), but are usually presented in terms of the electrical conductivity $\sig_{\mathrm{el}}$ instead, which makes sense only in a weakly-gauged description. Note that as $\rho$ and not $\sig_{\mathrm{el}}$ enters into this derivation, in principle this relation makes sense even when the electrodynamic sector is strongly coupled. Due to the issues with universality described earlier, we cannot claim that we have shown that this formula universally describes the decay rate. It is however in agreement with holographic results exhibited in \cite{Das:2022auy}, and it would be very interesting to compare it to recent lattice results \cite{figueroa1, Figueroa:2019jsi}, where it may be possible to independently measure $\rho$ and $\chi_A$ on the lattice. As described in those works, at the moment lattice computations for this prefactor are in disagreement with elementary hydrodynamic arguments based around a gauged vector current, and one might imagine that our more universal treatment is of value.

This expression suggests that as $\sB \to 0$ the relaxation rate $\Ga_{A}$ vanishes. At this point we should note that we have been working in a classical theory and have not included any sort of fluctuations, i.e. we have essentially interpreted the operator $\langle J\tilde{J}\rangle$ to be $\langle J \rangle \langle \tilde{J} \rangle$. This is clearly only an approximation, and it seems possible that incorporating fluctuations could result in a $\sB$-independent contribution to $\Ga_{A}$. This would then become the dominant effect at small $\sB$ \footnote{We thank L. Delacr\'{e}taz for illuminating discussions on this point.}. We leave investigation of this important point to later work; for now we note that in the presence of some limit (e.g. large $N$) allowing classicality it can in principle be separated from the considerations discussed here. 



We conclude with a brief symmary of our results. Had our symmetry simply be a product of explicitly broken 0-form $U(1)$, with a lifetime $(\Gamma_A)^{-1}$ and unbroken 1-form $U(1)$ symmetry, one would expect the theory in the deep IR at $t\gg (\Gamma_A)^{-1}$ to only consist of a 1-form symmetry, i.e. to be  indistinguishable from those in \cite{Grozdanov:2016tdf}. We show that the theory with a system with ABJ anomaly is differ from an ordinary MHD both in and out of equilibrium.

In the equilibrium sector, the physics that we get is that of the \textit{chiral separation effect} (CSE). Surprisingly, the formula that describes the CSE on our case exactly matches with the one present in Eq.(4.15) of \cite{Landsteiner:2016led}, where in the latter it was derived in a weakly-coupled way with non-dynamical electromagnetism. We see that if we correct for the electric and magnetic fields to $\mathcal{O}(k)$, then the functional form of the CSE remains the same. However, we have not presented the most general equilibrium action in the sense that it can have more terms in it and then one needs to check if the functional form of the CSE still remains same or if it receives correction (as shown in \cite{Jensen:2013vta}) even after addition of more terms to the equilibrium action. We shall return this exercise in future. However, the crucial point to note here is that we get this chiral MHD phenomenology even though we never make any reference to electromagnetism.

In the dissipative sector, we derive the physics of the \textit{chiral magnetic effect}. The coefficients appearing in this effect can be derived from Kubo formulae, as given in \cite{Grozdanov:2016tdf}, and in this sense they are somewhat universal. However, the dissipative action is not invariant under a diagonal shift symmetry. We shall return to this point later in future to resolve this issue with the dissipative action. We also found that, while the density in ordinary MHD relaxed to equilibrium through diffusion process with $\omega\sim -i q^2$, a theory with ABJ anomaly relaxed to the equilibrium configuration through subdiffusion process with $\omega \sim -i q^4$. While this is not the first time that such a mode is found, it would be interesting to investigate whether or not it is signature of non-invertible symmetry of this type.

\section*{Acknowledgements}

We would like to thank M. Baggioli, L. Delacr\'{e}taz, A. Donos, I. Garcia-Etxebarria, S. Grozdanov, P. Kovtun, N. Lohitsiri and T. Sulejmanpasic for insightful discussions. 
We would like to thank H. Liu and M. Landry for correspondence.
The work of NP and NI at Durham University is supported by STFC grant number ST/T000708/1.

\begin{appendix}

\section{Useful identities and conventions}
\subsection{Differential form identities} 
For completeness we record some identities involving differential forms:
\begin{align}
d(\om_p \wedge \eta_q) & = d\om_p \wedge \eta_q + (-1)^p \om_p \wedge d\eta_q \\
\om_p \wedge \eta_q & = (-1)^{pq} \eta_q \wedge \om_p \\
\om_p \wedge \star\eta_p & = \eta_p \wedge \star\om_p
\end{align} 
The square of the Hodge star acting on a $p$ form in $n$ dimensions on a metric with $s$ minus signs in its eigenvalues is
\be
\star^2 = (-1)^{s+p(n-p)}
\ee
\subsection{Discrete symmetries} \label{app:discrete} 
Here we record some background on the construction of Table \ref{CPT} of discrete symmetries. Note that $a^\mu$ is the axial source and hence it is a pseudo-vector and has the same transformation under discrete symmetry as that of $\overline{\Psi}\gamma^\mu\gamma^5\Psi$. Next let us look at $b_{\mu\nu}$. Since $b_{\mu\nu}$ is the source for the 2-form current, it couples in the action as $\epsilon^{\mu\nu}b_{\mu\nu}F_{\rho\sigma}$, where $F_{\rho\sigma}$ is the field strength of the vector gauge field. So, $b_{ij}$ couples to $F^{k\tau}$ and $B_k (=b_{k\tau})$ couples to $F^{ij}$. $F^{\mu\nu}$ being a 2-index object we will have the following transformations of its components under discrete symmetries, $F^{k\tau}$ will transform as $\overline{\Psi}\sigma^{k\tau}\Psi$ and $F^{ij}$ will transform as $\overline{\Psi}\sigma^{ij}\Psi$. Once we have identified the above discrete transformations, we may now use standard results (e.g. those in \cite{PeskinQFT:1995}).  

\section{Equilibrium configuration from gauging procedure} \label{sec:EqFromGauging}

In this section, we will analyse the equilibrium configuration from the point of view of gauging the the anomalous $U(1)$ symmetry. We will show that the gauging procedure put constraints on the equilibrium parameter of the ungauged theory. These kind of constaints are well-known but there are subtleties upon turning on the background fields gauge fields which play crucial roles in the perturbation around equilibrium configuration considered in the next section. 

In the simplest example of a theory with anomaly-free 0-form $U(1)$ global symmetry. The equilibrium partition function can be constructed in terms of thermodynamic quantities and background metric $g_{\mu\nu}$ and background gauge fields $a_\mu$ as in \cite{Banerjee:2012iz,Jensen:2012jh} namely\footnote{In the convention of \cite{Banerjee:2012iz,Jensen:2012jh}, the field strength is treated as first derivative quantity. In this paper, however, we will treat it as zeroth derivative quantity as in \cite{Kovtun:2016lfw}}
\begin{equation}
   W_0 =  - \log Z_0 =\int d^4x \sqrt{g} \left( p(T,g_{\mu\nu},\mu,f_{\mu\nu}) + \text{higher-derivative terms}\right)
\end{equation}
where $\mu/T = \log \left(\exp\left( \int^{1/T}_0 d\tau a_\tau\right)\right)$ is the $U(1)$ holonomy around the thermal cycle $\tau$ and $f =da$ is the field strength. From this, one can write the local expression for chemical potential as
\begin{equation}\label{eq:LocalChemPot}
    \mu = u^\mu\left( a_\mu + \partial_\mu \theta \right)
\end{equation}
where $u^\mu$ is the unit vector along the thermal $S^1$ direction which, for flat space, is nothing but $u^\mu = \delta^\mu_\tau$. While it is common to choose a gauge where $\mu = u^\mu a_\mu$, it is possible to turn have nonzero chemical potential without external gauge field $a_\tau$ by choosing the parameter $\theta = \mu \tau$ with a singularity at $\tau = 0\sim \beta$. This distinction is important as the chemical potential and background $a_\tau$ corresponds to different quantities when the $U(1)$ is the axial global symmetry. In this framework, the $U(1)$ current can be written as 
\begin{equation}
    j^\mu =\frac{1}{\sqrt{g}} \frac{\delta W}{\delta a_\mu} = \rho u^\mu + \nabla_\nu M^{\mu\nu}
\end{equation}
where $\rho = \partial p/\partial\mu$ and $M^{\mu\nu} = \partial p/\partial f_{\mu\nu}$. In this configuration there is no relation between $\rho$ and $M^{\mu\nu}$ except that they depends on the arbitrary function $p$.

The story is quite different upon promoting the background $a_\mu$ to the dynamical gauge field $A_\mu$. Upon doing this the partition function becomes 
\begin{equation}\label{eq:gaugedPartitionFunction}
-\log Z_\text{gauged}[b] = W_0 + W_{EM} + \int d^4x A_\mu j^\mu_\text{ext} \, , \qquad W_{EM} = \frac{1}{4 e^2} \int d^4x F_{\mu\nu} F^{\mu\nu}    
\end{equation}
with $F=dA$ and $\star j_{ext} =db$ is the external current. The equations of motion for $A_\mu$ implies that there is an additional relation 
\begin{equation}\label{eq:gauged-constraint}
    \frac{\delta}{\delta A_\mu} (W_0 + W_{EM}) + j^\mu_\text{ext} = 0 \, , \qquad \rho u^\mu + \nabla_\nu M^{\mu\nu}_\text{gauged} + j^\mu_\text{ext} = 0
\end{equation}
where $M^{\mu\nu}_\text{gauged} = \delta (W_0+W_{EM})/\delta F_{\mu\nu}$. We now see that there is a new relation between $\rho$ and $M^{\mu\nu}_\text{gauged}$ which does not exist before gauging. In a state where the electric field vanishes, this is nothing but the charge neutrality condition in plasma. It should also be noted that this form of the \eqref{eq:gauged-constraint} is applicable for arbitrarily nonlinear form of $W_{EM}$ which may or maynot be the Maxwell action. One may view this relation as a generalisation of Guass and Ampere's law.

\subsection{Gauging $U(1)$ symmetry in a theory with mixed anomaly}

We will now consider how a constraint such as \eqref{eq:gauged-constraint} is modified when the ungauged theory has a mixed anomaly.
We shall now consider a theory with mixed anomaly between vector $U(1)$ and axial $U(1)$ which yield the Ward identity \eqref{eq:WardIdenABJ} upon gauging. The ungauged theory partition function consist of 
\begin{equation}
    W_0 = W_\text{inv}[\mu_A, \mu_v, f_A,f_v] + W_\text{anom}[a,v, \mu_A,\mu_v,f_A,f_v]
\end{equation}
where $\mu_A,a,f_A=da$ and $\mu_v,v,f_v=dv$ are chemical potential, background gauge field and field strength of the axial and vector $U(1)$ global symmetry respectively.
This action is not invariant with respect to the background gauge transformation and so do the \textit{consistent} currents obtained via varying $W_0$ with respect to the the background gauge field. The invariant partition function $W_{cov}$ can be made out of $W_0$ by attaching the 4d theory to a 5d bulk with the Chern-Simons term $I_{CS}$ which satisfy $dI_{CS} = k f_A\wedge f_v\wedge f_v$, see e.g. \cite{Jensen:2013kka,Jensen:2013vta} for a modern summary. The \textit{covariant} currents can be obtained via the usual variation
\begin{equation}\label{eq:defjcov}
    j^\mu_{a,cov} = j^\mu_A + j^\mu_{A,BZ} \, , \qquad j^\mu_A = \frac{\delta W_0}{\delta a_\mu} \, ,\qquad j^\mu_{A,BZ} = \frac{\delta I_{CS}}{\delta a_\mu}
\end{equation}
and similar expression for $j^\mu_{v,cov}$ obtain by simply chaning $a \to v$. The expression for $j^\mu_{A,cov}$ and $j^\mu_{v,cov}$ is well-known in the literature in the scheme where $f_a,f_v$ are treated as first derivative derivative quantities. In our case, where we treat $f$ as a zeroth derivative quantity and focus on the case of space with $u^\mu =\delta^{\mu\tau}$ and $T$ be a constant, we find that 
\begin{equation}\label{eq:jcovjBZandM}
    j^i_{v,\text{cov}}= 2k \mu_a \mathcal{B}^i + \partial_j M^{ij}  \, , \qquad j^i_{v,BZ} = 2 k a_{t} \CB^i \, , \qquad M^{ij} =\epsilon^{ijk} \frac{\delta W_{inv}}{\delta \mathcal{B}_k}
\end{equation}
where $\mathcal{B}^i = \epsilon^{ijk} (f_v)_{jk}$ is the background magnetic field at this stage.

Upon gauging the vector $U(1)$ in 4d theory describes by $W_0$ i.e. promoting the background $v$ to the dynamical gauge field $V$ and add the source term and kinetic term as in \eqref{eq:gaugedPartitionFunction}, the equations of motion for $A_V$ implies that there is a condition on the consistent current 
\begin{equation}
  j^i_v + j^i_\text{ext} = 0
\end{equation}
We find that the spatial component becomes 
\begin{equation}
    \partial_j M^{ij}_\text{gauge} + 2k(\mu_A-a_{t}) \mathcal{B}^i + j^i_\text{ext} = 0
\end{equation}  
where $M^{ij}_\text{gauge} = \delta(W_\text{inv} + W_{EM})/\delta F_{V,ij}$ with $F_V =dV$ be the dynamical field strength. In the case where the partition function is dominated by magnetic field
\begin{equation}
    W_{inv}+W_{EM} = \int d^4x\,\left( \frac{1}{2\chi_B} \mathcal{B}_i\mathcal{B}^i + \text{subleading terms}\right)
\end{equation}
with $\chi_B$ be the susceptibility of the 1-form $U(1)$ density. In such a state, we find that, in the absence of the source.
\begin{equation}\label{eq:EqCond-MagDom}
    \epsilon^{ijk} \partial_j \mathcal{B}_k + 2\chi_B k(\mu_A-a_{t})\mathcal{B}^i = 0
\end{equation}
The relation \eqref{eq:EqCond-MagDom} must be satisfied for any theory with ABJ anomaly in a homogeneous configuration and it has very simple solutions.
Upon contracting with $\mathcal{B}_i$ one finds that this relation implies 
\begin{equation}
    \mu_A-a_{t} = 0 \, , \quad \mathcal{B}^i \ne 0\,\qquad\text{or}\qquad \mu_A \ne a_{t} \, ,
    \quad\mathcal{B}^i = 0\,.
\end{equation}
The choice where $\mu_A-a_{t}= 0$ is the only one that allows finite magnetic field and will be the equilibrium configuration of focus in the remaining of this work.

One may wonder, from the definition of chemical potential in \eqref{eq:LocalChemPot} why is it not possible for us to perform a background gauge transformation $a\to a + d\lambda$ to guarantee that $\mu$ is always $a_{t}$. There are at two ways to argue why is is not automatically satisfied. First of all, the current $j^\mu$ which coupled to $a$ is not a conserved current due to the r.h.s. of \eqref{eq:WardIdenABJ} and thus the source $a$ and $a + d\lambda$ are not equivalent. Second, one can see from a Landau-level calculation for Weyl fermion, see e.g. \cite{Landsteiner:2016led}. There, the source $a_{t}$ indicates the difference in energy at the tips of the right-handed and left-handed Weyl cone while the chemical potential $\mu$ is conjugate to the difference of occupation number between left- and right-handed fermions. 
\section{Defect operator insertion in equilibrium}\label{eqm_d_i}

In this section, we outline how the non-invertible defect leads to the Ward identity in \eqref{symms}. The non-invertible codimension-1 defect defined in \cite{Choi:2022jqy,Cordova:2022ieu} is on a closed surface $\Sigma$ can be written as
\be \label{eq:defDefectCLS}
 \hat{\CD}_{\frac{p}{N}}(\Sigma) = \exp\left[ i \int_\Sigma\left(\frac{2\pi p}{2N} \star j + \CA^{N,p}[\star_4 J/N]  \right) \right]
\ee
with $N$ and $p$ mod $N$ are two coprime and the 2-form $J$ is the conserved current associated to the 1-form global symmetry. $\CA^{N,p}$ is the Lagrangian density of the minimal TQFT with $\mathbb{Z}_N$ 1-form global symmetry with the 1-form $\mathbb{Z}_N$ anomaly parametrised by an integer $p$. In a particular case when $p=1$, we have 
\be
\exp\left[i\int_\Sigma \CA^{N,1}[\star_4 J/N]  \right] =\int D[\bar a] \exp\left[ i\int_\Sigma \left( \frac{N}{4\pi}\bar a\wedge d\bar a + \frac{1}{2\pi} \bar a\wedge \star_4 J \right) \right]
\ee
This describes the fractional quantum Hall system as integrating out $\bar a$ yield $N d\bar a + \star_4 J =0$ and resulting in Chern-Simons term with fractional coefficient $p/N$. The Chern-Simons living on the defect has a $\mathbb{Z}_N$ anomaly \cite{Gaiotto:2014kfa} which can be cancelled when attached to a bulk TQFT on $\CM$, such that $\d \CM= \Sigma$, with the following action \cite{Hsin:2018vcg}
\be
\CS_\text{bulk} = \int_{\CM} \left( \frac{N}{4\pi} B\wedge B + \frac{N}{2\pi} B\wedge dc \right)\label{eq:bulkSPT}
\ee
with $c$ is the 1-form $U(1)$ gauge field whose equation of motion forced $B$ to be the a $\mathbb{Z}_N$ 2-form gauge field (to be identified with $N B = \star_4 J$). The minimal TQFT $\CA^{N,p}$ is then defined as a theory living on $\Sigma=\d\CM$ with coefficient of $B\wedge B$ in \eqref{eq:bulkSPT} from $N/4\pi$ to $Np/4\pi$. The fusion algebra, which shows the non-invertibility $\hat{\CD}_{\frac{p}{N}} \times \hat{\CD}_{-\frac{p}{N}} \ne \mathds{1}$ can be found in \cite{Choi:2022jqy,Cordova:2022ieu}. An alternative description for $\hat{\CD}$ with the arguements $p/N$ extended to $U(1)$ valued can also be found in \cite{Karasik:2022kkq,GarciaEtxebarria:2022jky}.

A theory is said to have non-invertible global symmetry of this type when the operator $\hat{\CD}_{\frac{p}{N}}(\Sigma)$ is topological. That is, one can continuously deform the surface $\Sigma$ to $\Sigma'$ without changing the partition function. As a consequence, if we consider two nearby surface $\Sigma,\Sigma'$ which enclosed a small fluid element living in a small (and topologically trivial) $m_4$ such that $\d m_4 = \Sigma \cup \Sigma'$, we find that $\hat{\CD}(\Sigma)$ and $\hat{\CD}(\Sigma')$ are identical if only (consider $p=1$ for simplicity)
\be
\begin{aligned}\label{eq:1/NDefectCompt}
1 &=\exp\left[ \left(\int_\Sigma -\int_{\Sigma'}  \right)\left(\frac{2\pi }{2N} \star j + \CA^{N,1}[\star_4 J/N]  \right)\right]\\
&=\int D[\bar a] \exp\left[\int_{m_4} \left( \frac{\pi}{N} d\star j + \frac{N}{4\pi}d\bar a\wedge d\bar a + \frac{1}{2\pi}d\bar a \wedge \star_4 J  \right)\right] \\
&= \exp\left[ \int_{m_4} \left( \frac{\pi}{N} d\star j- \frac{1}{4\pi N} J\wedge J  \right) \right]
\end{aligned}
\ee
Converting the term in the parenthesis in components, we find the Ward identity \eqref{symms}.

To analyse the equilibrium of a theory with non-invertible symmetry, we can put it on a manifold $S^1\times \mathbb{R}^3$ as in 
Section \ref{eqm}. In this case, let us consider a local fluid element in $m_4$ which also contain the thermal cycle $S^1$. The topological condition \eqref{eq:1/NDefectCompt} implies the (dimensionally reduced) Ward identity \eqref{eq:WardDimReduced}\footnote{
It is possible that, upon dimensionally reduced on the thermal $S^1$, the non-invertible defect in \eqref{eq:defDefectCLS} will give rises to codimension-0 and codimension-1 defect in three dimensions similar to those in \cite{Damia:2022rxw}. This is an interesting future direction, however we will only consider the consequence of \eqref{eq:1/NDefectCompt} at the level of the Ward identity in this work.
}. Notice that, had the equilibrium partition function only consist of susceptibility of 0-form and 1-form global symmetry and thus described by the action
\be
\CS =  \int d^3x \left[ \frac{1}{2}\chi_A (a_\tau)^2 + \frac{1}{2}\chi_B (B_i B^i) \right]\, .
\ee
i.e. when $\chi_\CO$ in \eqref{second_a2} is turned off, the topological property \eqref{eq:1/NDefectCompt} is trivially satisfied. This is because both $j^i$ and $\epsilon_{ijk} J^{i\tau} J^{jk}$ vanishes identically. However, as one turned on $\chi_\CO$ in \eqref{second_a2}, then $j^i=0$ but $\epsilon_{ijk} J^{i\tau} J^{jk} =2\chi_B^i \d_i f \ne 0$ which means that the non-invertible defect is not topological (see \eqref{eq:constiReln-Equi} for the notation). Thus, the action with nonzero $\chi_A,\chi_B$ and $\chi_\CO$ has to be modified in a nontrivial way as demonstrated in Section \ref{sec:EquiCompat}.


\section{Defect operator insertion in dissipative theory}
\label{app:defectop} 

\subsection{Defect operator insertion} 
In this section we shall discuss defect insertions in the dissipative action given in \ref{diss_actn}. Following \cite{Liu:2018kfw} we have, $(1,2)$ as the two degrees of freedom in {\it Schwinger-Keldysh} formalism. Also, in the $r-a$ basis, the $``r"$ type fields are somewhat like the physical fields and the $``a"$ type fields are somewhat like noise. So, to go from here to the equilibrium phase we neglect the time derivatives and put $\phi_a = 0$. The basic transformation among the two bases is,
\begin{align}
    &\phi_r = \frac{1}{2}\left(\phi_1 + \phi_2\right), \qquad \phi_a = \left(\phi_1 - \phi_2\right), \label{r-a} \\
    &\phi_1 = \phi_r + \frac{1}{2}\phi_a, \qquad \phi_2 = \phi_r - \frac{1}{2}\phi_a. \label{1-2}
\end{align}

\subsubsection{Dissipative Action}
Let us consider the dissipative acion in the main text, i.e. \eqref{diss_actn}
\begin{align}
    \CL[a,b;\theta,\Phi;\Sigma,C] = \CL_{\text{MHD}} + \CL_a -2\left(\Sigma_a\wedge dC_r + \Sigma_r\wedge dC_a\right) + \CL_{\text{anom}}[\theta,C]. \label{diss_actn_NI}
\end{align}
where $a,b$ denote external sources, $\theta, \Phi$ denote dynamical fields and $\Sigma, C$ are auxiliary fields or Lagrange multipliers. 
The 1-form currents as follows,
\begin{align}
    &j_r = \frac{\delta \CS}{\delta a_a}, \qquad j_a = \frac{\delta \CS}{\delta a_r}, \label{r-a1form} 
\end{align}
where $\CS$ is now to be understood as the dissipative action.
Similarly, we obtain the 2-from currents as,
\begin{align}
    &J_r = \frac{\delta \CL}{\delta \Sigma_a} = dC_r\, ,  \qquad J_a = \frac{\delta \CL}{\delta \Sigma_a} = dC_a\, . 
\end{align}
This implies in terms of these new $C_r$ and $C_a$ fields, the currents $J_r$ and $J_a$ are now identically conserved.

The part of the action that involves the axial charge fluctuation can be written in the $1,2$ basis as follows
\begin{align}
    \CL_a &= \frac{i\sigma}{\beta}A^2_{ai} + \chi_A A_{a0}B_{r0} - \sigma A_{ai}A_{ri,0}, \label{r-a_La} \\
    &= \frac{i\sigma}{\beta}\left[A_{1i}^2 + A_{2i}^2 - 2A_{1i}A_{2i}\right] + \frac{\chi_A}{2}\left[A_{1t}^2 - A_{2t}^2\right] \nonumber\\
    &-\frac{\sigma}{2}\left[A_{1i}\left(A_{1i,t} + A_{2i,t}\right) - A_{2i}\left(A_{1i,t} + A_{2i,t}\right)\right], \label{1-2_La}
\end{align}
where $A = a+d\theta$. Similar decomposition can also be done in for the MHD part i.e.
\begin{align}
    \CL_{\text{MHD}} = \frac{i\rho}{\beta_0}\tilde G^2_{aij} + \chi_B \tilde G_{a0i}\tilde G_{r0i} - \sigma \tilde G_{aij}\tilde G_{rij,0}. \label{L_MHD} 
\end{align}
where $\tilde G=b+d\Phi+\Sigma$, as well as the Lagrange multiplier
\begin{align}
    \Sigma_a\wedge dC_r + \Sigma_r\wedge dC_a = \Sigma_1\wedge dC_1 - \Sigma_2\wedge dC_2\, . \label{r-a1-2sig}
\end{align}
For this action to be compatible with the non-invertible defect, we have to add additional terms $\CL_\text{anom}$ of the following form
\begin{align}
    \CL_{\text{anom}} = -4K\left(\theta_1 dC_1\wedge dC_1 - \theta_2 dC_2\wedge dC_2\right) \label{anom_1-2}\,.
\end{align}
At this stage, $K$ can be any function of thermodynamic quantities which may or may not has to do with the constant $k=1/16\pi^2$ in the Ward identity. Here, we will show that, for the defect insertion to be consistent, the function $K$ must be a constant and equal to $k$. 

\subsubsection{Non-invertible defect operator insertion}
Due to the doubling of the degrees of freedom we now have two defect operators constructed as in \cite{Choi:2022jqy,Cordova:2022ieu}. Inserting the non-invertible defect turns the Schwingker-Keldysh generating function into 
\be
Z = \exp\left(-i\CS) \right)\to Z' = \hat{\CD}_1 \hat{\CD}_2  \exp\left(i \CS \right)
\ee 
where $\hat{\CD}_1$ and $\hat{\CD}_2$ are 
\begin{align}
    &\hat{\CD}_1 = \int D[\bar a_1] \exp\left(\int_{\CM} \left(\frac{2\pi}{2N} \star j_1 + \frac{N}{4\pi} {\Bar a}_1\wedge d{\Bar a}_1 + \frac{1}{2\pi} {\Bar a}_1\wedge dC_1\right)\right), \label{def1} \\
    &\hat{\CD}_2 = \int D[\bar a_2]\exp\left(\int_{\CM} \left(\frac{2\pi}{2N} \star j_2 + \frac{N}{4\pi} {\Bar a}_2\wedge d{\Bar a}_2 + \frac{1}{2\pi} {\Bar a}_2\wedge dC_2\right)\right)\,, \label{def2}
\end{align}    
where if $\CM = \mathbb{R}^3$ then defect is inserted at $t=0$ (temporal insertion) and if $\CM = \mathbb{R}^{1,2}$ then defect is inserted at $z=0$ (spatial insertion). 

We shall assume that, apriori, to begin with, all fields are smooth across the defects. Let us first consider inserting the non-invertible defect at $z=0$. The currents involves in this analysis are,
\begin{equation}\label{eq:jrzjaz}
\begin{aligned}
    &j_{rz} = \frac{\delta \CS}{\delta a_{az}} = \frac{2i\sigma}{\beta}A_{az} - \sigma\d_t A_{rz},  \\
    &j_{az} = \frac{\delta \CS}{\delta a_{rz}} = \sigma\d_t A_{az},
\end{aligned}
\end{equation}
or in the 1,2 basis, we have 
\be
\begin{aligned}
\star &j_{1z} = j_{rz} + \frac{1}{2}j_{az} = \frac{2i\sigma}{\beta}\left(A_{1z} - A_{2z}\right) - \sigma\d_t\left(A_{2z}\right),  \\
    \star &j_{2z} = j_{rz} - \frac{1}{2}j_{az} = \frac{2i\sigma}{\beta}\left(A_{1z} - A_{2z}\right) - \sigma\d_t\left(A_{1z}\right). 
\end{aligned}
\ee

Consider the equation of motion of $\theta_1,\theta_2$ in the presence of the non-invertible defect, we get
\be
4K \, dC_s\wedge dC_s + \frac{2i\sigma}{\beta}\left[\theta_1-\theta_2\right]_{,zz} + \frac{2i\sigma}{\beta}\left(\frac{2\pi}{2N}\right)\frac{d}{dz}\delta(z) + (\ldots) = 0\,, \label{th1_eom_sp}
\ee
where $s=1,2$. Here $(\ldots)$ includes terms with less than two $z$ derivatives. Both equations yield the solution 
\be
\Delta(\theta_1-\theta_2)\equiv (\theta_1-\theta_2)\Big\vert_{z+\epsilon} - (\theta_1-\theta_2)\Big\vert_{z-\epsilon} = -\frac{2\pi}{2N}
\ee
The equation of motion for $\bar a_1,\bar a_2$, we have 
\begin{align}
    &\left.Nd{\Bar a}_s + dC_s\right|_{z=0} = 0, 
\end{align}
which can be used to replaced $d\bar a_s$ in terms of $dC_s$. Finally, $C_1$'s and $C_2$'s equations of motion are
\begin{align}
    &2d\Sigma_s + 8Kd\left(\theta_s dC_s\right) + (-1)^{s}\frac{d{\Bar a}_s}{2\pi}\delta(z) = 0, \label{C1_eom_sp} 
\end{align}
Combined all the equations of motion together, one finds that, 
\begin{align}
    &K=\frac{1}{16\pi^2}, \label{1_sp}
\end{align}
where note that, when $s=1$ and $s=2$, \eqref{C1_eom_sp} is satisfied by the following conditions 
\be
\Delta\theta_1=-\frac{\pi}{N}\, , \qquad \Delta\theta_2=0\,, \qquad \text{and}\qquad \Delta\theta_1=0\, , \qquad \Delta\theta_2=\frac{\pi}{N}\,,
\ee
respectively.

So, in order for the theory to be compatible with non-invertible defect insertion we see that $K=k$. 

Similar analysis can be done for the defect insertion localised in the time direction at $t=0$, with $j^z_{r,a}$ in \eqref{eq:jrzjaz} replaced by $j^t_{r,a}$, and results in \eqref{1_sp} without giving additional constraints.

\end{appendix} 
\bibliographystyle{utphys}

\bibliography{all}

\end{document}